\documentclass[11pt]{article}

\usepackage[DIV13]{typearea}
\usepackage{amsmath} 
\usepackage{amssymb}
\usepackage{amsfonts}
\usepackage{amscd}
\usepackage{graphicx}
\usepackage{epsfig}
\usepackage[footnotesize,bf]{caption2}
\usepackage{cite}
\usepackage{bbold}
\usepackage{longtable}
\usepackage[center,footnotesize,hang]{subfigure}
\usepackage{url}

\newcommand{\Eqref}[1]{Eq.\eqref{#1}}

\newcommand{\Tabref}[1]{Table \ref{#1}}

\newcommand{\Secref}[1]{Section \ref{#1}}

\newcommand{\Rep}[1]{\underline{\mbox{\textbf{#1}}}}
\newcommand{\MoreRep}[2]{\underline{\mbox{\textbf{#1}}} _{\mbox{\textbf{#2}}}}
\newcommand{\Groupname}[2]{$ {#1} _{#2} $}

\newcommand{\eps}{\epsilon}

\begin{document}

\begin{titlepage}
\begin{flushright}
SISSA  45/2010/EP
\end{flushright}
\vspace*{5mm}

\begin{center}
{\Large\sffamily\bfseries
\mathversion{bold} $\mu-\tau$ Symmetry and Charged Lepton Mass Hierarchy
in a Supersymmetric \Groupname{D}{4} Model 
\mathversion{normal}}
\\[13mm]
{\large
C.~Hagedorn~\footnote{E-mail: \texttt{hagedorn@sissa.it}}
and
R.~Ziegler~\footnote{E-mail: \texttt{ziegler@sissa.it}}} 
\\[5mm]
{\small \textit{
SISSA, Scuola Internazionale Superiore di Studi Avanzati\\
and\\
INFN-Sezione di Trieste\\
via Bonomea 265, I-34136 Trieste, Italy 
}}
\vspace*{1.0cm}
\end{center}
\normalsize
\begin{abstract}
\noindent In this note we discuss a supersymmetric (SUSY) $D_4 \times Z_5$ model 
which leads to vanishing reactor mixing angle $\theta_{13}=0$ and maximal
atmospheric mixing $\theta_{23}=\pi/4$ in the lepton sector at leading order (LO),
due to the preservation of non-trivial distinct $D_4$ subgroups in the charged lepton
and neutrino sectors, respectively. The solar mixing angle $\theta_{12}$ remains undetermined and
is expected to be of order one.
Since right-handed charged leptons transform as singlets under $D_4$,
 the charged lepton mass hierarchy can be naturally accounted for. 
The model predicts inverted mass hierarchy for neutrinos.
 Additionally, we show that, unlike in most of the other models of this type, all vacuum expectation 
values (VEVs) of gauge singlets (flavons) can be fixed through mass parameters of the superpotential.
Next-to-leading order (NLO) corrections to lepton masses and mixings are calculated and shown to be under control, 
especially the corrections to $\theta_{23}=\pi/4$ and $\theta_{13}=0$ are of the order of the 
generic expansion parameter $\eps \approx 0.04$ and arise dominantly from the charged
lepton sector.
\end{abstract}

\end{titlepage}

\setcounter{footnote}{0}

\section{Introduction}
\label{sec:intro}

The observed mixing pattern in the lepton sector \cite{nudata},
\begin{equation}
 \!\!\!\!\!\!\!\!
\sin ^{2} \theta_{12} = 0.318 ^{+0.042} _{-0.028} \; , \;\;
\sin ^{2} \theta_{23} = 0.50 ^{+0.13} _{-0.11} \;\;\; \mbox{and} \;\;\;
\sin ^{2} \theta_{13} = 0.013 ^{+0.026} _{-0.013} \;\;\; (2 \, \sigma) \; ,
\end{equation}
is well compatible with special mixing patterns in which $\theta_{13}$ vanishes and $\theta_{23}$ is 
maximal,
\begin{equation}
\label{eq:th23th13fixed}
\sin ^2 \theta_{23} = \frac{1}{2} \;\;\; \mbox{and} \;\;\; \sin \theta_{13} = 0 \; . 
\end{equation}
\Eqref{eq:th23th13fixed} is derived from a $\mu-\tau$ symmetric neutrino mass matrix \cite{mutau}, in the charged
lepton mass basis. At the same time the solar mixing angle $\theta_{12}$ is not predicted. 
It is tempting to assume a symmetry to be the origin of such a mixing pattern.

It has been shown \cite{dntheory} that in case of a non-trivial breaking of a dihedral flavor symmetry
in the lepton (quark) sector one element $U_{\alpha\beta}$ of the lepton (quark) mixing matrix is given in terms of 
group theoretical quantities only
\footnote{For specific realizations see \cite{D4SUSY_us,thetaC,GL1,GLS3,symmetrygeneral}.}
\begin{equation}
|U_{\alpha\beta}|= \left| \cos \left( \frac{\pi \, (k_{1} - k_{2}) \, \mathrm{j}}{n} \right) \right|
\end{equation}
where $n$ refers to the group index of the dihedral group $D_n$, $\rm j$ to the index of the two-dimensional
irreducible representation $\MoreRep{2}{j}$ under which two of the three left-handed fields transform
and $k_{1,2}$ are associated to the generating elements $\mathrm{B A}^{k_1}$ and $\mathrm{B A}^{k_2}$ 
of the $Z_2$ subgroups preserved in the charged lepton (down quark) and neutrino (up quark) sectors, respectively. ($\rm A$ and $\rm B$
are the generators of the original dihedral group $D_n$.) 
Note that $k_1$ and $k_2$ have to be distinct to get a non-trivial value for $|U_{\alpha\beta}|$. Considering the lepton sector,
 one sees that choosing $n=4$ makes it possible
to achieve e.g. $|U_{\mu3}|=\frac{1}{\sqrt{2}}$. Using a specific set of flavons to break $D_4$ to the subgroup $Z_2$ generated
by $\mathrm{B A}^{k_1}$
 in the charged lepton sector, also $|U_{\tau3}|=\frac{1}{\sqrt{2}}$ and $U_{e3}=0$ can be enforced, thus leading to $\mu-\tau$ symmetric lepton mixings.

In the model presented here $D_4$ is accompanied by the cyclic symmetry $Z_5$. Both symmetries are only spontaneously
broken by flavon VEVs. We use as framework the Minimal SUSY Standard Model (MSSM). 
It is not the first time that $\mu-\tau$ symmetry is deduced from the group $D_4$ \cite{GL1,Ishimori_D4,Ishimori_D4_SUSY,D4SUSY_us}. However,
in these models producing the mass hierarchy among charged leptons is usually non-trivial. 
In contrast, the observed charged lepton mass hierarchy
\begin{equation}
\label{eq:masshierch}
m_e: m_\mu: m_\tau \approx \epsilon^2: \epsilon: 1 \;\;\; \mbox{with} \;\;\; \epsilon \approx \lambda^2 \approx 0.04 \, ,
\end{equation}
with $\lambda$ being the Wolfenstein parameter \cite{Wolfpara}, is a natural outcome of our model. 
In order to achieve this, it is essential that the right-handed charged
lepton fields transform as (different) singlets under the flavor symmetry $D_4$, instead of being in the representations $\Rep{1}+\Rep{2}$.
Notice that also the introduction of a $U(1)$ symmetry is not necessary for ending up with \Eqref{eq:masshierch}, see as well
\cite{S3alt,Yin,AM}.

At LO, our model leads to $\theta_{13}=0$ and $\theta_{23}=\pi/4$ with no constraints on the solar mixing angle $\theta_{12}$,
apart from predicting its value generically to be of order one. In the limit of preserved subgroups $m_\mu$ and $m_\tau$ are generated of the
correct order of magnitude. Since $m_\tau$ arises from a non-renormalizable operator small and moderate values of $\tan\beta$
are preferred. Neutrinos exhibit inverted mass hierarchy and the lightest neutrino mass $m_3$ fulfills $m_3 \gtrsim 0.015$ eV. 
We find at LO a strong correlation among $m_3$ and the measure of neutrinoless double beta decay $|m_{ee}|$, 
a rather weak one among $m_3$ and $\tan\theta_{12}$ as well
as a restricted range for Majorana phases. These results are very similar to those found in \cite{D4SUSY_us}.
We study the model to NLO and show that corrections to the
predictions $\theta_{13}=0$ and $\theta_{23}$ being maximal are of order $\eps$ and dominated by the charged lepton sector.
Also $\theta_{12}$ undergoes corrections; but these are not particularly interesting, because its value is not a prediction of our model anyway.
The mass of the electron is generated by NLO corrections. The masses of the other charged leptons and neutrinos are slightly corrected
as well.

Furthermore, an appropriate construction of the flavon superpotential allows us to 
fix all flavon VEVs through couplings with positive
mass dimension. By choosing the latter to be of order $\eps \Lambda$, where $\Lambda$ is the generic cutoff scale of our theory, all flavon VEVs
are of order $\eps\Lambda$. Fixing all flavon VEVs by mass parameters is usually 
 not the case in models of such type which thus
encounter free parameters among the flavon VEVs (equivalent to flat directions in the flavon (super)potential), see e.g. 
\cite{S3alt,D4SUSY_us,thetaC,AF2}.

The paper is structured as follows: in \Secref{sec:grouptheory} we briefly introduce the group
 $D_4$ and discuss its subgroups. The model is outlined and the LO results for lepton
masses and mixings are presented in \Secref{sec:outline}. \Secref{sec:flavons} contains the discussion of the flavon
superpotential at LO and NLO. In \Secref{sec:fermions_NLO}
 NLO corrections to lepton masses and mixings are shown to be well under control.
We conclude in \Secref{sec:conclusions}.

\mathversion{bold}
\section{$D_4$ Group Theory}
\mathversion{normal}
\label{sec:grouptheory}

The dihedral group $D_4$ has eight (distinct) elements and five irreducible representations denoted here as
$\MoreRep{1}{i}$, $\rm i=1,...,4$ 
and $\Rep{2}$. All representations are real and only $\Rep{2}$
is faithful. $D_4$ is generated by $\rm A$ and $\rm B$ which can be chosen as \cite{dngrouptheory}
\begin{equation}
\label{eq:generators}
\rm A =\left(\begin{array}{cc} 
                            i & 0 \\
                            0 & -i 
          \end{array}\right) \;\;\; \mbox{and} \;\;\; \rm 
		B=\left(\begin{array}{cc} 
                                       0 & 1 \\
                                       1 & 0 
                  \end{array}\right) 
\end{equation}
for $\Rep{2}$. Note that we have chosen $\rm A$ to be a complex matrix, although $\Rep{2}$ is a real
representation. For $(a_1,a_2)^t \sim \Rep{2}$ we thus find that $(a_2^*, a_{1}^*)^t$
transforms as $\Rep{2}$ under $D_4$. The generators $\rm A$ and $\rm B$ of the one-dimensional
representations are
\begin{eqnarray}
\label{eq:generators_1}
\MoreRep{1}{1} \;\;\; &:& \;\;\; \mathrm{A}=1 \; , \;\; \mathrm{B}=1 \, ,\\
\MoreRep{1}{2} \;\;\; &:& \;\;\; \mathrm{A}=1 \; , \;\; \mathrm{B}=-1 \, ,\\
\MoreRep{1}{3} \;\;\; &:& \;\;\; \mathrm{A}=-1 \; , \;\; \mathrm{B}=1 \, ,\\
\MoreRep{1}{4} \;\;\; &:& \;\;\; \mathrm{A}=-1 \; , \;\; \mathrm{B}=-1 \, .
\end{eqnarray}
The character table can be found in e.g. \cite{D4SUSY_us}.
$\rm A$ and $\rm B$ fulfill the relations
\begin{equation}\label{eq:genrelations}
\mathrm{A}^{4} =\mathbb{1} \;\;\; , \;\;\; \rm B^2=\mathbb{1} \;\;\; \mbox{and}
\;\;\; \rm ABA=B \; . 
\end{equation}
The Kronecker products involving $\MoreRep{1}{i}$ are the following
\begin{equation}\nonumber
\MoreRep{1}{i} \times \MoreRep{1}{i}= \MoreRep{1}{1} \; , \;\; 
\MoreRep{1}{1} \times \MoreRep{1}{i}= \MoreRep{1}{i} \;\; \mbox{for} \;\; \rm i=1,...,4 
\; , \;\;
\MoreRep{1}{2} \times \MoreRep{1}{3}= \MoreRep{1}{4} \; , \;\;
\MoreRep{1}{2} \times \MoreRep{1}{4}= \MoreRep{1}{3} \;\; \mbox{and} \;\;
\MoreRep{1}{3} \times \MoreRep{1}{4}= \MoreRep{1}{2} \; .
\end{equation}
For $s_i \sim \MoreRep{1}{i}$ and $(a_1,a_2)^t \sim \Rep{2}$ we find
\begin{equation}\nonumber
\left( \begin{array}{c} s_1 a_1 \\ s_1 a_2
\end{array} \right) \sim \Rep{2} \;\; , \;\;\;
\left( \begin{array}{c} s_2 a_1 \\ -s_2 a_2
\end{array} \right) \sim \Rep{2} \;\; , \;\;\;
\left( \begin{array}{c} s_3 a_2 \\ s_3 a_1
\end{array} \right) \sim \Rep{2} \;\;\; \mbox{and} \;\;\;
\left( \begin{array}{c} s_4 a_2 \\ -s_4 a_1
\end{array} \right) \sim \Rep{2} \;\; .
\end{equation}
The four one-dimensional representations contained in $\Rep{2} \times \Rep{2}$ 
 read for $(a_1,a_2)^t$, $(b_1, b_2)^t$ $\sim \Rep{2}$
\begin{equation}\nonumber
a_1 b_2 + a_2 b_1 \sim \MoreRep{1}{1} \;\; , \;\;\;
a_1 b_2 - a_2 b_1 \sim \MoreRep{1}{2} \;\; , \;\;\;
a_1 b_1 + a_2 b_2 \sim \MoreRep{1}{3} \;\;\; \mbox{and} \;\;\;
a_1 b_1 - a_2 b_2 \sim \MoreRep{1}{4} \;\; .
\end{equation}
These formulae are special cases of the expressions, given for dihedral symmetries $D_n$ with general index $n$, 
which can be found e.g. in \cite{dntheory,kronprods}.

In order to understand how maximal atmospheric mixing and vanishing $\theta_{13}$ arise in our model,
 it is relevant to discuss the subgroups of $D_4$.
All its subgroups are abelian: $Z_2 \cong D_1$, $Z_4$ and $D_2 \cong Z_2 \times Z_2$.
In the following we are only interested in $Z_2$ subgroups generated by $\mathrm{B} \, \mathrm{A}^k$
with $k=0,...,3$, because these are preserved in the charged lepton and neutrino sectors at LO. Noting  
\begin{equation}\nonumber
(\mathrm{B} \, \mathrm{A}^{k})^{2} = \mathrm{B} \, \mathrm{A}^{k} \mathrm{B} \, \mathrm{A}^{k}
= \mathrm{B} \, \mathrm{A}^{k-1} \mathrm{B} \, \mathrm{A}^{k-1} = \dots 
= \mathrm{B}^{2} = \mathbb{1}
\end{equation}
shows that $\mathrm{B} \, \mathrm{A}^{k}$ indeed generates
a $Z_2$ symmetry. Since $k$ can take integer values between $0$ and $3$, we find four possible
$Z_2$ subgroups of this type. Apart from VEVs of fields transforming trivially under $D_4$, a $Z_2$ group given through 
$\mathrm{B} \mathrm{A}^{k}$ is left unbroken by a non-vanishing VEV of a singlet transforming as $\MoreRep{1}{3}$ if $k=0,2$  
holds, and of one singlet transforming as $\MoreRep{1}{4}$ for $k=1,3$. Additionally, it is left intact by fields $\psi_{1,2}$ 
forming a doublet, if their VEVs have the following structure
\begin{equation}
\left( \begin{array}{c} \langle \psi_{1} \rangle \\
 \langle \psi_{2} \rangle \end{array} \right) \propto \left(
\begin{array}{c} \mathrm{e}^{-\frac{\pi i k}{2}}\\ 1
\end{array}
\right) \; .
\end{equation}
As mentioned in the Introduction, in order to get non-trivial mixing
the $Z_2$ subgroups preserved in the charged lepton and the neutrino sectors have to have different indices
$k_l$ and $k_\nu$. To achieve maximal atmospheric mixing, we need e.g. $k_l$ to be odd,
whereas $k_\nu$ has to be even.
This is analogous to the constraints found for the indices $k_{l,\nu}$ in \cite{D4SUSY_us}.
In the following section we show that it is phenomenologically
irrelevant whether $k_l$ is $1$ or $3$ as well as whether $k_\nu$ is $0$ or $2$, i.e. the result
for the lepton mixing angles is in all cases $\theta_{13}=0$ and $\theta_{23}=\pi/4$.
In \Secref{sec:flavons} a simple flavon superpotential is constructed which naturally leads to an odd index 
$k_l$ and an even index $k_\nu$. 

\section{Outline of the Model and Results at LO}
\label{sec:outline}

In this section the transformation properties of all fields (apart from the driving fields responsible for the vacuum
alignment) under the
flavor symmetry $D_4 \times Z_5$ are presented. Similar to the model discussed in 
\cite{D4SUSY_us} the framework is the MSSM and the left-handed lepton doublets $L_i$ transform as 
$\MoreRep{1}{1} + \Rep{2}$ under $D_4$. In order to accommodate the charged lepton mass hierarchy,
 we assign the right-handed charged leptons $e^c_i$ to the three singlets
$\MoreRep{1}{2} + \MoreRep{1}{3} + \MoreRep{1}{4}$ and not to $\MoreRep{1}{1} + \Rep{2}$ as done in \cite{D4SUSY_us}.
In contrast to the model given in \cite{D4SUSY_us} we assign the $Z_5$ charges to the matter superfields and 
$h_{u,d}$ such that $h_{u,d}$ are uncharged under $D_4 \times Z_5$ and, more importantly, we allow the
 generations of one type, $L_i$ and $e^c_i$, to transform differently under $Z_5$.
In this way the additional cyclic symmetry $Z_5$ does not only play the role of a symmetry which separates 
the charged lepton and neutrino sectors (at LO),
 but also plays the role of a Froggatt-Nielsen symmetry \cite{FN}. As one can see,
we use for the five different multiplets $L_1$, $L_D=(L_2, L_3)^t$, $e^c_i$ all five different possible $Z_5$ charges.
As in \cite{D4SUSY_us} the light neutrino masses arise from the effective operator $L_i L_j h_u^2/ \Lambda$. 
 The flavons relevant for achieving $\mu-\tau$ symmetry in the lepton sector
at LO, are $\psi_e$ and $\psi_\nu$ which form doublets under $D_4$. The VEV of $\psi_e$ is aligned through the
flavon superpotential in such a way that either the $Z_2$ subgroup generated 
by $\mathrm{B} \mathrm{A}$ or by $\mathrm{B} \mathrm{A}^3$ is preserved in the
charged lepton sector, whereas the special form of $\langle\psi_{\nu;1,2}\rangle$ preserves either a $Z_2$ subgroup arising
from $\mathrm{B}$ or from $\mathrm{B} \mathrm{A}^2$. As we will show it is not relevant for phenomenology which of the two subgroups
in each sector is chosen. The relevant aspect is the fact that different subgroups are preserved. 
 The two additional flavons $\eta_1$ and $\eta_3$ 
are Froggatt-Nielsen type fields. Since they do not transform under $D_4$ they do not
play a role in the preservation of different subgroups of $D_4$. 
They allow for non-zero $(11)$ and $(23)$, $(32)$ entries in the neutrino mass matrix at LO
and are relevant for generating the appropriate hierarchy among the charged lepton masses, because 
$m_\tau$ arises dominantly from $L_D e^c_3 h_d \psi_e/\Lambda$, while the mass of
the muon is generated through the operator $L_D e^c_2 h_d \psi_e \eta_1/\Lambda^2$. The mass of the electron originates mainly
from one subleading operator involving three flavons, $L_1 e^c_1 h_d \psi_e \psi_\nu \eta_1/\Lambda^3$.
 Since the tau lepton mass stems from a non-renormalizable
operator, $v_d = \langle h_d \rangle$ is expected to be of the order of the electroweak scale and thus small and moderate values of
$\tan\beta= \langle h_u\rangle/ \langle h_d \rangle = v_u/v_d$ are preferred in this model. The right order of magnitude of
$m_\tau$ can be achieved for $\langle \psi_e \rangle/\Lambda \approx \epsilon \approx 0.04$. The ratio $m_\mu:m_\tau \sim \epsilon:1$
requires that also $\langle\eta_1\rangle/\Lambda \approx \epsilon$. As derived from the flavon superpotential all
flavon VEVs can be of order $\epsilon \Lambda$ and thus the electron mass $m_e$ is expected to fulfill $m_e:m_\tau \sim \epsilon^2:1$
due to its origin from a three flavon operator. Apart from generating the electron mass, operators involving
several flavons (as well as shifts in the flavon VEVs) lead to small deviations from the LO result that $\theta_{23}$
is maximal and $\theta_{13}$ vanishes. As we show below, these deviations are dominated by corrections associated
to the charged lepton sector, while all subleading effects in the neutrino sector are of relative order $\eps^2$.
 All fields appearing in the Yukawa couplings and their transformation properties under $D_4 \times Z_5$ 
are captured in \Tabref{tab:particles}.
\begin{table}
\begin{center}
\begin{tabular}{|c||c|c|c|c|c||c|c||c|c||c|c|}\hline
Field & $L_{1}$ & $L_D$ & $e^{c}_{1}$ & 
$e^{c}_{2}$ & $e^{c}_{3}$ & $h_{u}$ & $h_{d}$
& $\psi_{e; 1,2}$ & $\psi_{\nu; 1,2}$ & $\eta_1$ & $\eta_3$\\ 
\hline
$D_4$ & $\MoreRep{1}{1}$ & 
$\Rep{2}$ & $\MoreRep{1}{2}$ & $\MoreRep{1}{3}$ & $\MoreRep{1}{4}$ & $\MoreRep{1}{1}$ 
 & $\MoreRep{1}{1}$ & $\Rep{2}$ & $\Rep{2}$ & $\MoreRep{1}{1}$ & $\MoreRep{1}{1}$\\
$Z_5$ & $\omega^2$ & $\omega$ & 
$1$ & $\omega^3$ & $\omega^4$ & $1$ & $1$ & $1$
& $\omega^2$ & $\omega$ & $\omega^3$\\ 
\hline
\end{tabular}
\end{center}
\begin{center}
\begin{minipage}[t]{12cm}
\caption[]{Particle content of the model. $L_i$ denotes the three left-handed lepton $SU(2)_L$ doublets,
$e^c_i$ the right-handed charged leptons and $h_{u,d}$ the MSSM Higgs doublets.
The flavons $\psi_{e; 1,2}$ and $\psi_{\nu; 1,2}$ only transform
under $D_4 \times Z_5$. Also the Froggatt-Nielsen type fields $\eta_1$ and $\eta_3$
are gauge singlets and only carry a non-zero $Z_5$ charge.  
$\omega=\mathrm{e}^{\frac{2 \pi i}{5}}$ is the generating element of $Z_5$.
\label{tab:particles}}
\end{minipage}
\end{center}
\end{table}

In order to elucidate how maximal atmospheric mixing and vanishing reactor mixing angle arise, we first take
into account only operators which are suppressed by at maximum $1/\Lambda^2$ and in which the flavon doublet $\psi_\nu$ only couples to
neutrinos, whereas $\psi_e$ couples to charged leptons. 
In this way the $Z_2$ subgroups in charged lepton and neutrino
sectors remain exactly preserved. We show subsequently that the inclusion
of further operators, compatible with the symmetries of the model, and shifts in the flavon VEVs only slightly correct the results achieved at this level.
With the above restrictions the allowed operators in the superpotential are 
\footnote{We do not list the operator $L_1 e^c_2 h_d \psi^2_e/\Lambda^2$, because plugging in the vacuum given in \Eqref{eq:vac_LO}
renders this possible contribution to the charged lepton mass matrix zero. However, we discuss the operator in Section 5,
since it leads to a non-vanishing contribution, once the shifts in the flavon VEVs are taken into account.}
\begin{eqnarray}
\label{eq:wlatLO}
W_l = & & \frac{y_1^e}{\Lambda} \left( L_2 e^c_3 h_d \psi_{e; 1} - L_3 e^c_3 h_d \psi_{e; 2}  \right)
        + \frac{y_2^e}{\Lambda^2} \left( L_2 e^c_2 h_d \psi_{e; 1} + L_3 e^c_2 h_d \psi_{e; 2} \right) \eta_1\\
    &+& \frac{y_1^\nu}{\Lambda^2} L_1 L_1 h_u^2 \eta_1 
	+ \frac{y_2^\nu}{\Lambda^2} (L_2 L_3 + L_3 L_2) h_u^2 \eta_3 
        + \frac{y_3^\nu}{\Lambda^2} L_1 (L_2 h_u^2 \psi_{\nu; 2} + L_3 h_u^2 \psi_{\nu; 1})
\nonumber
\\  &+&	\frac{y_3^\nu}{\Lambda^2} (L_2 \psi_{\nu; 2} + L_3 \psi_{\nu; 1}) h_u^2 L_1 
         \, . \nonumber 
\end{eqnarray}
As shown in \Secref{sec:flavons}, the vacuum structure is
\begin{equation}
\label{eq:vac_LO}
\langle \psi_{e; 2} \rangle = - i \, \rho_e \, \langle \psi_{e; 1} \rangle = w_e  \, , \;\;
\langle \psi_{\nu; 2} \rangle = \rho_\nu \, \langle \psi_{\nu; 1} \rangle = w_\nu  \, , \;\;
\langle \eta_1 \rangle = w_1 \;\;\; \mbox{and} \;\;\; \langle \eta_3 \rangle = w_3 
\end{equation}
with $\rho_{e,\nu} = \pm 1$. The different choices of $\rho_{e,\nu}$ correspond to the different possible values of
the subgroup index $k_l$ and $k_\nu$, respectively. Since they are not uniquely fixed by the
flavon superpotential, we keep $\rho_{e,\nu}=\pm 1$ as parameters and 
check that all possibilities lead to $\theta_{13}=0$ and $\theta_{23}=\pi/4$.
We arrive at fermion mass matrices of the form ($M_l$ given in left-right convention)
\begin{equation}
\label{eq:fermionsatLO}
M_{l} =  
\left( 
\begin{array}{ccc}  
0 & 0 & 0 \\ 
0  & i \,\rho_e \, y_2^e \, w_1/\Lambda  &  i \,\rho_e \, y_1^e \\ 
0 &  y_2^e \, w_1/\Lambda & -y_1^e 
\end{array} 
\right) \, \frac{w_e}{\Lambda} \, v_d \;\;\; \mbox{and} \;\;\;
M_{\nu} =  
\left( 
\begin{array}{ccc}  
y_1^\nu w_1 & y_3^\nu w_\nu & \rho_\nu y_3^\nu w_\nu \\ 
y_3^\nu w_\nu & 0 &  y_2^\nu w_3 \\ 
\rho_\nu y_3^\nu w_\nu &  y_2^\nu w_3 & 0 
\end{array} 
\right) \, \frac{v_u^2}{\Lambda^2} \; .
\end{equation}
For 
\begin{equation}
\label{eq:sizeVEV}
w_e, \, w_\nu, \, w_1 , \, w_3 \approx \epsilon \, \Lambda
\end{equation}
with $\epsilon \approx 0.04$, muon and tau lepton masses read
\begin{equation}
\label{eq:mls_LO}
m_\mu = \sqrt{2} |y^e_2| w_1 w_e v_d/\Lambda^2 \approx \epsilon^2 v_d\;\;\; \mbox{and} \;\;\; m_\tau = \sqrt{2} |y^e_1| w_e v_d/\Lambda 
\approx \epsilon v_d\, ,
\end{equation}
while the electron remains massless at this stage and acquires a mass from operators with three flavon insertions, see \Eqref{eq:me_mass}.
The charged lepton mass matrix can be diagonalized through the usual biunitary transformation ($U_l$ and $V_l$) so that
\begin{equation}
U_l^\dagger M_l V_l = \mbox{diag} (m_e,m_\mu,m_\tau)
\end{equation}
with $U_l$ given by
\begin{equation}
\label{eq:Ul_LO}
U_{l} =  
\left( 
\begin{array}{ccc}  
1 & 0 & 0 \\ 
0 & i \rho_e/\sqrt{2} & -i \rho_e/\sqrt{2}\\ 
0 & 1/\sqrt{2} & 1/\sqrt{2} 
\end{array} 
\right) 
\end{equation}
and $V_l$ is a diagonal matrix, ensuring that the entries of $M_l$ in the charged lepton mass basis are real and positive.
The unitary matrix $U_\nu$ diagonalizing $M_\nu M_\nu^\dagger$ can be written in the form
\begin{equation}
\label{eq:Unu_LO}
U_{\nu} =  
\left( 
\begin{array}{ccc}  
\cos \phi_\nu & \sin \phi_\nu \, \mathrm{e}^{-i\gamma_\nu}& 0 \\ 
-\rho_\nu \, \sin \phi_\nu \, \mathrm{e}^{i\gamma_\nu}/\sqrt{2} & \rho_\nu \, \cos \phi_\nu/\sqrt{2} & -\rho_\nu/\sqrt{2}\\ 
-\sin \phi_\nu \, \mathrm{e}^{i\gamma_\nu}/\sqrt{2} & \cos \phi_\nu/\sqrt{2} & 1/\sqrt{2} 
\end{array} 
\right) 
\end{equation}
with $\phi_\nu$ and $\gamma_\nu$ being real functions of the entries of $M_\nu$. 
For the absolute values of the lepton mixing matrix we then find
\begin{equation}
|U_{MNS}| = 
\left( 
\begin{array}{ccc}  
|\cos \phi_\nu| & |\sin \phi_\nu| & 0 \\ 
|\sin \phi_\nu|/\sqrt{2} & |\cos \phi_\nu|/\sqrt{2} & 1/\sqrt{2}\\ 
|\sin \phi_\nu|/\sqrt{2} & |\cos \phi_\nu|/\sqrt{2} & 1/\sqrt{2}
\end{array} 
\right)  
\end{equation}
showing that $\theta_{23}=\pi/4$ and $\theta_{13}=0$ hold, while the solar mixing angle $\theta_{12}$ is not fixed in our model and related
to the angle $\phi_\nu$. As one can see the form of $|U_{MNS}|$ and thus the result for the mixing angles do not depend on a particular choice of 
$\rho_{e}$ and $\rho_\nu$.

In the charged lepton mass basis the neutrino mass matrix $M_\nu^\prime=U_l^\dagger M_\nu U_l^*$ reveals a texture zero in its (23)
entry.
According to the findings in \cite{texturezero} a consequence of the presence of this texture zero 
together with $\theta_{13}=0$ is that the neutrino mass spectrum has to have inverted hierarchy and
furthermore the lightest neutrino mass $m_3$ cannot vanish. This is in our model not 
a generic result attributed to the flavor symmetry $D_4 \times Z_5$ and its breaking pattern, but is due to the
fact that a flavon coupling to the neutrino sector and transforming as ($\MoreRep{1}{3}$, $\omega^3$)
under ($D_4$,$Z_5$) is absent. 
 A numerical analysis similar to the one found in \cite{D4SUSY_us} can also be performed for this model. The
results of such an analysis are quantitatively very similar to those found in \cite{D4SUSY_us} for the case of a non-zero
(23) entry of the neutrino mass matrix $M'_\nu$ in a spontaneously CP violating framework. These include that
the measure of neutrinoless double beta decay $|m_{ee}|$ fulfills $|m_{ee}| \approx m_3$, a weak correlation between $\tan \theta_{12}$
and $m_3$ and rather strong correlations of the two Majorana phases and the mass $m_3$ as well as a lower bound on $m_3$ around $0.015$ eV.
Plots showing these results can be found in \cite{D4SUSY_us} in Figures 1 to 4.
All results are found to be independent of the different choices of $\rho_\nu$ and $\rho_e$.
Noting that the third neutrino mass $m_3$ is given
by $|y^\nu_2| w_3 v_u^2/\Lambda^2$, we find that in order to correctly reproduce the light neutrino mass scale of around $0.1$ eV 
a value of the cutoff scale $\Lambda \approx 4 \times 10^{12}$ GeV for $v_u\approx 100$ GeV is necessary.

Comparing our results to those found in \cite{D4SUSY_us} we see that the difference of the two models lies in the fact that in \cite{D4SUSY_us}
the subgroup in the charged lepton sector is $D_2$, while it is only a $Z_2$ subgroup in our model. As a consequence, the model in \cite{D4SUSY_us}
leads to $\mu-\tau$ symmetric mixing through preserving certain $D_4$ subgroups only, whereas in the model here the absence of a flavon transforming
as $\MoreRep{1}{4}$ under $D_4$ from the charged lepton sector is relevant for achieving $\theta_{23}=\pi/4$ and $\theta_{13}=0$, see for
details Appendix B in \cite{D4SUSY_us}.

\section{Flavon Superpotential}
\label{sec:flavons}

In the following we align the VEVs of the fields $\psi_e$ and $\psi_\nu$
correctly and fix all VEVs through mass parameters of the superpotential, thus avoiding flat
directions in the potential associated to VEVs which remain undetermined. We consider
the $F$-terms of a new set of fields, the driving fields, as source of the alignment conditions.
These fields are like the flavons gauge singlets and transform in general under $D_4\times Z_5$.
We assume the existence of a continuous $R$-symmetry $U(1)_R$ under which all matter superfields
have charge $+1$, flavons and the superfields $h_{u,d}$ charge $0$
and driving fields charge $+2$, so that the superpotential, responsible for the alignment of the flavon VEVs, will
be linear in the driving fields.

\subsection{Renormalizable Level}
\label{subsec:flavons_LO}

In order to align the vacuum of $\psi_\nu$ we introduce one driving field,
$\sigma^{0}_{\nu}$, transforming as $\MoreRep{1}{4}$ under $D_4$,
similar to \cite{D4SUSY_us}, and as $\omega$ under $Z_5$. The alignment of the vacuum
of $\psi_e$ is achieved in a similar way through coupling it to a field $\sigma^0_e$
transforming as $\MoreRep{1}{3}$ and being invariant under $Z_5$. We add a field
$\chi^0_e$ which is neutral under $D_4 \times Z_5$ to allow for a coupling of
mass dimension two which fixes the size of the VEV of $\psi_e$. The VEVs of $\psi_\nu$
and of the Froggatt-Nielsen fields $\eta_1$ and $\eta_3$ are deduced from the 
$F$-terms of three further fields, $\chi^0_\nu$, $\eta^0_3$ and $\eta^0_4$ transforming
as shown in \Tabref{tab:driving}. Note that also the field $\eta^0_4$ allows for a 
coupling with positive mass dimension in the flavon superpotential. Apart from $\sigma^0_e$
and $\sigma^0_\nu$, responsible for the alignment of $\langle \psi_{e;1,2}\rangle$ and $\langle \psi_{\nu;1,2}\rangle$,
all other driving fields transform trivially under $D_4$. 
The flavon superpotential $W_{fl}$ is given, at the renormalizable level, as
\footnote{We can safely neglect the term $\chi^0_e h_u h_d$, because the VEVs of all flavons are much larger than the 
electroweak scale.}
\begin{eqnarray}
\label{eq:wfLO}
W_{fl} = & & a_e  \, \sigma^0_e \left( \psi_{e; 1}^2 + \psi_{e; 2}^2 \right) + M_1^2 \, \chi^0_e + b_e \, \chi^0_e \psi_{e; 1} \psi_{e; 2}
\\ \nonumber
  & +&  a_\nu  \, \sigma^0_\nu \left( \psi_{\nu; 1}^2 - \psi_{\nu; 2}^2 \right)  
  + b_\nu \, \chi^0_{\nu} \, \psi_{\nu; 1} \, \psi_{\nu; 2}  +  c_\nu \, \chi^0_{\nu}\eta_1\eta_3 
\\ \nonumber
  &+ &  d \, \eta^0_3 \left( \psi_{e; 1}\psi_{\nu; 2} + \psi_{e; 2}\psi_{\nu; 1} \right) + f \, \eta^0_3 \eta_1^2
 + M_2 \, \eta^0_4 \eta_1 + g \, \eta^0_4 \eta_3^2 
\end{eqnarray}
\begin{table}
\begin{center}
\begin{tabular}{|c||c|c|c|c|c|c|}\hline
Field & $\sigma^0_{e}$ & $\sigma^0_{\nu}$ & $\chi^0_{e}$ & $\chi^0_{\nu}$ & $\eta^0_3$ & $\eta^0_4$ \\ 
\hline
$D_4$ & $\MoreRep{1}{3}$ & $\MoreRep{1}{4}$ & $\MoreRep{1}{1}$ & $\MoreRep{1}{1}$ & $\MoreRep{1}{1}$ & $\MoreRep{1}{1}$\\
$Z_5$ & $1$ & $\omega$ & $1$ & $\omega$ & $\omega^3$ & $\omega^4$ \\ 
\hline
\end{tabular}
\end{center}
\begin{center}
\begin{minipage}[t]{12cm}
\caption[]{Driving fields of the model necessary to align the VEVs of $\psi_e$ and $\psi_\nu$ and to
fix the flavon VEVs through mass parameters of the superpotential. All these fields have charge $+2$
under $U(1)_R$.
\label{tab:driving}}
\end{minipage}
\end{center}
\end{table}
with $a_e$, $b_e$, $a_\nu$, $b_\nu$, $c_\nu$, $d$, $f$ and $g$ being complex numbers with absolute values of order one.
Assuming that the flavons acquire their VEVs in the SUSY limit, i.e. (soft) SUSY breaking effects can be safely neglected, we can use the 
$F$-terms of the driving fields to
determine the vacuum structure of the flavons. The first two equations read
\begin{equation}
\label{eq:FtermsLO_1}
\frac{\partial W_{fl}}{\partial \sigma^0_e}= a_e  \, \left( \psi_{e; 1}^2 + \psi_{e; 2}^2 \right) = 0 
\;\;\; \mbox{and} \;\;\;
\frac{\partial W_{fl}}{\partial \sigma^0_\nu}= a_\nu \, \left( \psi_{\nu; 1}^2 - \psi_{\nu; 2}^2 \right) = 0 \; .
\end{equation}
We find as solutions
\begin{equation}
\langle \psi_{e;2} \rangle = - i \rho_e \langle \psi_{e;1} \rangle = w_e 
\;\;\; \mbox{and} \;\;\;
\langle \psi_{\nu;2} \rangle = \rho_\nu \langle \psi_{\nu;1} \rangle = w_\nu 
\end{equation}
with $\rho_e = \pm 1$ and $\rho_\nu = \pm 1$. 
We see that we preserve a $Z_2$ subgroup of $D_4$ generated by either $\mathrm{B \, A}$ ($k_l=1$) or $\mathrm{B \, A}^3$ ($k_l=3$) 
in the charged lepton sector, whereas in the neutrino sector the subgroup is either generated by $\rm B$ ($k_\nu=0$)
or by $\mathrm{B \, A}^2$ ($k_\nu=2$). From 
\begin{equation}
\label{eq:FtermsLO_2}
\frac{\partial W_{fl}}{\partial \chi^0_e}= M_1^2 + b_e \, \psi_{e; 1} \psi_{e; 2} = 0
\end{equation}
it follows that the VEV of $\psi_{e}$ is fixed through the mass scale $M_1$ to be 
\begin{equation}
w_e ^2 = i \rho_e \frac{M_1^2}{b_e} \; .
\end{equation}
The last three $F$-term equations,
\begin{subequations}
\label{eq:FtermsLO_3}
\begin{eqnarray}
\frac{\partial W_{fl}}{\partial \chi^0_\nu}&=& b_\nu  \, \psi_{\nu; 1} \psi_{\nu; 2} + c_\nu \, \eta_1 \eta_3 = 0 \; ,\\
\frac{\partial W_{fl}}{\partial \eta^0_3}&=& d \, \left( \psi_{e; 1}\psi_{\nu; 2} + \psi_{e; 2}\psi_{\nu; 1} \right) 
+  f \, \eta_1^2 = 0 \; , \\ 
\frac{\partial W_{fl}}{\partial \eta^0_4}&=& M_2 \, \eta_1 + g \, \eta_3^2 = 0 \; ,
\end{eqnarray}
\end{subequations}
allow us to determine the VEVs of $\psi_{\nu}$, $\eta_1$ and $\eta_3$ as functions of the two mass parameters $M_1$ and $M_2$.
For $M_1$ and $M_2$ being of the order $\epsilon \Lambda$ we find that all flavon VEVs are of that order as well. $\langle \psi_e \rangle \neq 0$
is a necessary consequence of \Eqref{eq:FtermsLO_2} showing that $D_4$ is always spontaneously broken. The other flavon VEVs
could in principle vanish. However, if we assume that the VEV of one of the fields $\psi_\nu$, $\eta_1$ and $\eta_3$ does not
vanish then the non-vanishing of the other two ones follows. 
Due to the fact that all terms in $W_{fl}$ are by construction linear
in the driving fields the $F$-terms of the flavons vanish in any case for vanishing VEVs of all driving fields.
Then also the allowed term $\chi^0_e h_u h_d$ does not give rise to a $\mu$-term.

\subsection{Corrections from Non-Renormalizable Terms}
\label{subsec:flavons_NLO}

Including non-renormalizable terms with three flavons leads to corrections of the alignment achieved at LO.
We find nine terms to contribute at this level to the flavon superpotential
(again all possible terms involving the superfields $h_u$ and $h_d$ are neglected)
\begin{eqnarray}
\label{eq:Deltawfl}
\Delta W_{fl}  = & & x_1 \, \sigma^0_e \eta_1 \left( \psi_{\nu; 1}^2 + \psi_{\nu; 2}^2 \right)/\Lambda
+ x_2 \, \sigma^0_e \eta_3 \left( \psi_{e; 1} \psi_{\nu; 1} +  \psi_{e; 2} \psi_{\nu; 2} \right)/\Lambda
\\ \nonumber 
&+& x_3 \, \chi^0_e \eta_1 \psi_{\nu; 1} \psi_{\nu; 2}/\Lambda 
+ x_4 \, \chi^0_e \eta_3 \left( \psi_{e; 1} \psi_{\nu; 2} +  \psi_{e; 2} \psi_{\nu; 1} \right)/\Lambda
\\ \nonumber
&+&  x_5 \, \chi^0_e \eta_1^2 \eta_3/\Lambda +  x_6 \, \chi^0_\nu \eta_3^3/\Lambda
\\ \nonumber
&+& x_7 \, \eta^0_3 \eta_3 \psi_{\nu; 1} \psi_{\nu; 2}/\Lambda
+ x_8 \, \eta^0_3 \eta_1 \eta_3^2 /\Lambda
+ x_9 \, \eta^0_4 \eta_1 \psi_{e; 1} \psi_{e; 2}/\Lambda \; .
\end{eqnarray}
Note that there is no correction at this level involving the driving field $\sigma^0_\nu$. 
We can parametrize the VEVs of $\psi_e$ and $\psi_\nu$ as
\begin{equation}
\langle \psi_e \rangle = \left( \begin{array}{c}
i \rho_e (w_e + \delta w_{e;1}) \\ w_e + \delta w_{e;2}
\end{array} \right) \;\;\; \mbox{and} \;\;\;
\langle \psi_\nu \rangle = \left( \begin{array}{c}
 \rho_\nu (w_\nu + \delta w_{\nu;1}) \\ w_\nu + \delta w_{\nu;2}
\end{array}\right) 
\end{equation}
and find for the shifts in linear expansion that
\begin{eqnarray}
\label{eq:VEVshifts}
&& \frac{\delta w_{e;i}}{w_e} \, , \, \frac{\delta w_{\nu;i}}{w_\nu} \sim \epsilon \, , \\
\nonumber
&& \!\!\!\!\!\!\!\!\!\!\!\!\!\!\!\!\!\!\!\!\!\!\!\!
\delta w_{e;1} \neq  \delta w_{e;2} \;\;\; \mbox{and} \;\;\; \delta w_{\nu;1} =  \delta w_{\nu;2} \, .
\end{eqnarray}
Since the shifts of $\langle \psi_{\nu;i} \rangle$ are the same, the vacuum alignment is preserved up to this level
and we can absorb the shifts $\delta w_{\nu;i}$ into a redefinition of $w_\nu$. The equality of $\delta w_{\nu;i}$
is due to the fact that at the first non-renormalizable level the superpotential $\Delta W_{fl}$ does not contain
terms involving $\sigma^0_\nu$.
The shifts of the VEVs of the singlets $\eta_1$ and $\eta_3$ are of the same order of 
magnitude as $\delta w_{e;i}$ and $\delta w_{\nu;i}$. 
However, we do not mention them explicitly, because their effect on lepton masses and mixings can always be absorbed 
into a redefinition of Yukawa couplings or VEVs $w_1$ and $w_3$. 

\section{Lepton Masses and Mixings at NLO}
\label{sec:fermions_NLO}

In general, NLO corrections arise from two sources: $(i)$ shifts in the flavon VEVs
and $(ii)$ operators with two and more flavon insertions, evaluated by plugging in the LO form of the VEVs, 
which have not been considered in \Secref{sec:outline}. As we
show in the following, all such additional contributions change the LO results only slightly. However, their discussion
is relevant, because such terms generate the electron mass and govern the deviations from the LO results $\theta_{23}=\pi/4$ 
and $\theta_{13}=0$. We discuss all such terms which give
rise to contributions up to and including $\epsilon^3$ (in units of $v_d$ and $v_u^2/\Lambda$, respectively)
as well as for the (13) element of the charged lepton 
mass matrix also corrections of order $\eps^4 v_d$.

Considering the charged lepton sector we find the following additional operators involving two and more flavons
which contribute to the (23) and (33) elements of $M_l$
(order one coefficients are omitted and operators are not written in $D_4$ components)
\begin{equation}
\frac{1}{\Lambda^2} \, L_D e^c_3 h_d \psi_\nu \eta_3 + \frac{1}{\Lambda^3} \, L_D e^c_3 h_d \psi_e^3
\end{equation}
where the last term actually gives rise to two independent contributions.
In a similar way the (22) and (32) elements get corrected through
\begin{equation}
\frac{1}{\Lambda^3} \, L_D e^c_2 h_d \psi_\nu^3 + \frac{1}{\Lambda^3} \, L_D e^c_2 h_d \psi_e \eta_3^2
+ \frac{1}{\Lambda^3} \, L_D e^c_2 h_d \psi_\nu \eta_1 \eta_3
\end{equation}
with the first term being responsible for two independent contributions.
At the same time these entries also receive contributions from  
 the shifts in the VEV of $\psi_{e}$ which we indicate by
\footnote{As mentioned in \Secref{subsec:flavons_NLO} the shift in the VEV of the field $\eta_1$
can be absorbed into the Yukawa couplings or the LO VEV itself and thus its contribution is not displayed. The same holds for $\eta_3$.}
\begin{equation}
\frac{1}{\Lambda} \, L_D e^c_3 h_d \delta \psi_e + \frac{1}{\Lambda^2} \, L_D e^c_2 h_d \delta \psi_e \eta_1 \; . 
\end{equation}
The (21) and (31) elements are generated only through three flavon insertions of the form
\begin{equation}
\frac{1}{\Lambda^3} \, L_D e^c_1 h_d \psi_e \psi_\nu^2
+ \frac{1}{\Lambda^3} \, L_D e^c_1 h_d \psi_\nu \eta_1^2
+ \frac{1}{\Lambda^3} \, L_D e^c_1 h_d \psi_e \eta_1 \eta_3 \, .
\end{equation}
Note that the first operator gives rise to three independent terms. 
Similarly, the (11) element arises from
\begin{equation}
\label{eq:me_mass}
\frac{1}{\Lambda^3} \, L_1 e^c_1 h_d \psi_e \psi_\nu \eta_1
\; .
\end{equation}
 The generation of the (12) and (13) elements is somewhat special, because the lowest order
operators which could give rise to non-zero (12) and (13) elements are 
\begin{equation}
\frac{1}{\Lambda^2} \, L_1 e^c_2 h_d \psi_e^2
\;\;\; \mbox{and} \;\;\;
\frac{1}{\Lambda^2} \, L_1 e^c_3 h_d \psi_\nu^2
\, .
\end{equation}
However, plugging in the LO result for the VEVs of $\psi_e$ and $\psi_\nu$ we see that these operators
give no contribution. Taking into account the
shifts arising at a relative order $\epsilon$ we find that then the (12) element is generated, whereas
the (13) element still vanishes. Operators with three flavons also contribute to the (12) element at
this level
\begin{equation}
\frac{1}{\Lambda^3} \, L_1 e^c_2 h_d \psi_\nu^2 \eta_1
+ \frac{1}{\Lambda^3} \, L_1 e^c_2 h_d \psi_e \psi_\nu \eta_3 \; .
\end{equation}
The (13) element only originates from operators involving four flavons
\begin{equation}
\frac{1}{\Lambda^4} \, L_1 e^c_3 h_d \psi_e^2 \eta_1 \eta_3
+ \frac{1}{\Lambda^4} \, L_1 e^c_3 h_d \psi_e \psi_\nu \eta_1^2
+ \frac{1}{\Lambda^4} \, L_1 e^c_3 h_d \psi_e^2 \psi_\nu^2 \; .
\end{equation}
Again, the last operator allows for two independent contractions.
At this level we expect that the (13) element also receives a contribution from subleading shifts in the VEVs of $\psi_\nu$
which we however do not calculate, because the specific form of the (13) element is not relevant for the 
analysis of lepton masses and mixings. Thus, we can parametrize the charged lepton mass matrix including NLO
corrections as
\begin{equation}
\label{eq:Ml_NLO}
M_{l} =  
\left( 
\begin{array}{ccc}  
\beta^e_1 \eps^2 & \beta^e_2 \eps^2 & \beta^e_3 \eps^3\\ 
\beta^e_4 \eps^2  & i\rho_e\alpha^e_2 \eps+\beta^e_6\eps^2&  i\rho_e\alpha^e_1 +\beta^e_7\eps \\ 
\beta^e_5 \eps^2  & \alpha^e_2 \eps & -\alpha^e_1
\end{array} 
\right) \, \eps \, v_d
\end{equation}
taking into account the sizes of flavon VEVs and of their shifts as given in \Eqref{eq:sizeVEV} and \Eqref{eq:VEVshifts}. 
Through re-phasing of right-handed fields we can make $\alpha^e_1$, $\alpha^e_2$ and $\beta^e_5$ real and positive.
The other parameters, apart from $\eps$, are in general complex numbers with absolute values of order one.
Note further that the parameters $\alpha^e_{1,2}$ are up to corrections of order $\eps$  
determined by the LO operators given in \Eqref{eq:wlatLO}.

We find as result for the charged lepton masses
\begin{equation}
\label{eq:mls_NLO}
m_e = \left( |\beta^e_1| \eps^3 + \mathcal{O}(\eps^4) \right) \, v_d \; , \;\;
m_\mu = \left( \sqrt{2} \alpha^e_2 \eps^2 + \mathcal{O}(\eps^3) \right) \, v_d 
\;\;\; \mbox{and} \;\;\;
m_\tau = \left( \sqrt{2} \alpha^e_1 \eps + \mathcal{O}(\eps^2) \right) \, v_d 
\end{equation}
and thus confirm earlier statements about the size and origin of the electron mass in our model.
The unitary transformation applied to the left-handed charged leptons in order to diagonalize $M_l$
is, up to the first correction in $\eps$ in each matrix element, of the form
\begin{equation}
\label{eq:Ul_NLO}
U_{l} \approx  
\left( 
\begin{array}{ccc}  
1 - \left(\frac{|\beta^e_2|}{2 \alpha^e_2}\right)^2 \eps^2 & \frac{\beta^e_2}{\sqrt{2} \alpha^e_2} \eps
                    & -\frac{\beta^e_3}{\sqrt{2} \alpha^e_1} \eps^3 \\ 
-i \rho_e \frac{\beta^{e\,*}_2}{2 \alpha^e_2} \eps 
& \frac{1}{\sqrt{2}} \left( i \rho_e + \frac{\beta^{e\,*}_7}{2\alpha^e_1} \eps \right)& -\frac{1}{\sqrt{2}}\left(i \rho_e +\frac{\beta^e_7}{2\alpha^e_1}\eps \right)\\ 
-\frac{\beta^{e \, *}_2}{2\alpha^e_2} \eps & \frac{1}{\sqrt{2}}\left(1 + i\rho_e \frac{\beta^{e\,*}_7}{2\alpha^e_1}\eps\right) 
& \frac{1}{\sqrt{2}}\left( 1+ i\rho_e \frac{\beta^e_7}{2\alpha^e_1}\eps\right) 
\end{array} 
\right)  \; .
\end{equation}

We analyze the corrections to the neutrino mass matrix in a similar way and find that the following operators
 contribute to the (11) element
\begin{equation}
\frac{1}{\Lambda^3} \, L_1 L_1 h_u^2 \eta_3^2
+ \frac{1}{\Lambda^4} \, L_1 L_1 h_u^2 \psi_e^2 \eta_1
\end{equation}
and that corrections to the (23) element come from
\begin{equation}
\frac{1}{\Lambda^4} \, L_D L_D h_u^2 \eta_1^3 +
\frac{1}{\Lambda^4} \, (L_D L_D)_1 h_u^2 (\psi^2_e \eta_3)_1 +
\frac{1}{\Lambda^4} \, (L_D L_D)_1 h_u^2 (\psi_e \psi_\nu \eta_1)_1
\; .
\end{equation}
Here we denote by $(\cdots)_1$ the contraction to a trivial singlet $\MoreRep{1}{1}$ of $D_4$.
All these contributions can be absorbed into the LO result. The relation between the (12) and (13) element is disturbed at a
relative order $\epsilon^2$ through the terms
\begin{equation}
\frac{1}{\Lambda^4} \, L_1 L_D h_u^2 \psi_e^2 \psi_\nu
+ \frac{1}{\Lambda^4} \, L_1 L_D h_u^2 \psi_e \eta_1^2 \; .
\end{equation}
Note that the first operator generates three independent contributions.
Since the shift of the VEVs of $\psi_\nu$ is at lowest order aligned with the LO result, 
see \Eqref{eq:VEVshifts}, we do not encounter
any correction to the LO relation between the (12) and (13) elements at the relative level of $\epsilon$. Among the corrections
of order $\epsilon^3$ (in units of $v_u^2/\Lambda$) we also expect corrections due to possible deviations from 
$\langle \psi_{\nu;1}\rangle = \langle \psi_{\nu;2}\rangle$ at the level of $\epsilon^3 \Lambda$.
The (22) and (33) elements of $M_\nu$ receive two types of corrections. The first one still preserves
the $\mu-\tau$ symmetric structure of the LO result and generates entries of order $\epsilon^3 v_u^2/\Lambda$
\begin{equation}
\frac{1}{\Lambda^4} \, (L_D L_D)_3 h_u^2 (\psi_e^2 \eta_3)_3
+ \frac{1}{\Lambda^4} \, (L_D L_D)_3 h_u^2 (\psi_e \psi_\nu \eta_1)_3 \, .
\end{equation}
The second type breaks $\mu-\tau$ symmetry and stems from very similar operators
\begin{equation}
\frac{1}{\Lambda^4} \, (L_D L_D)_4 h_u^2 (\psi_e^2 \eta_3)_4
+ \frac{1}{\Lambda^4} \, (L_D L_D)_4 h_u^2 (\psi_e \psi_\nu \eta_1)_4 \, .
\end{equation}
As one can see the only difference is the contraction to another $D_4$ singlet; in the first case we
contract to a $\MoreRep{1}{3}$, indicated through $(\cdots)_3$, whereas in the second case the contraction
is to $\MoreRep{1}{4}$, denoted by $(\cdots)_4$. Also these $\mu-\tau$ symmetry breaking contributions 
arise at the level $\epsilon^3 v_u^2/\Lambda$.
Thus, at NLO $M_\nu$ takes the form
\begin{equation}
M_{\nu} =  
\left( 
\begin{array}{ccc}  
\alpha^\nu_1 & \alpha^\nu_3 + \beta^\nu_1 \eps^2& \rho_\nu \alpha^\nu_3\\ 
\alpha^\nu_3 + \beta^\nu_1 \eps^2 & \beta^\nu_2 \eps^2 &  \alpha^\nu_2\\ 
\rho_\nu \alpha^\nu_3 &  \alpha^\nu_2 & \beta^\nu_3 \eps^2 
\end{array} 
\right) \, \eps \, \frac{v_u^2}{\Lambda} \; .
\end{equation}
Note that the parameters $\alpha^\nu_{1,2,3}$ and $\beta^\nu_{1,2,3}$ are complex numbers with absolute values of order one.
Similar to the charged lepton sector, the parameters $\alpha^\nu_{1,2,3}$ coincide at LO in the expansion in $\eps$
with those given in \Eqref{eq:wlatLO} and \Eqref{eq:fermionsatLO}.

Recomputing the mass spectrum of the neutrinos we find that all masses are corrected by terms of relative order $\eps^2$.
Relevant for calculating the deviations from maximal atmospheric mixing and $\theta_{13}=0$ is the form of the
eigenvector associated to the third neutrino mass $m_3$. We find that its LO form, see \Eqref{eq:Unu_LO}, receives
corrections only at order $\eps^2$. Thus, the results for the mixing angles $\theta_{13}$ and $\theta_{23}$ at NLO are 
dominated by corrections coming from the charged lepton sector, see \Eqref{eq:Ul_NLO}, and we get
\begin{eqnarray}
&& |U_{e3}| =\sin\theta_{13} \approx \frac{|\beta^e_2|}{2\alpha^e_2} \eps \; ,\\
&& \sin^2\theta_{23} \approx \frac{1}{2} \left( 1 - \rho_\nu \frac{\mathrm{Re}(\beta^e_7)}{\alpha^e_1} \eps\right) \; .
\end{eqnarray}
The solar mixing angle $\theta_{12}$ whose exact value is not predicted in our model, but only given in terms of the parameters
of the neutrino mass matrix at LO, undergoes also corrections of order $\eps$. However, these are not of particular interest.

\section{Conclusions}
\label{sec:conclusions}

We have presented a SUSY $D_4$ model which leads to maximal atmospheric mixing and vanishing $\theta_{13}$, while keeping the solar 
mixing angle $\theta_{12}$ undetermined, but expected to be in general large. These predictions originate from the non-trivial breaking of $D_4$
to distinct $Z_2$ subgroups generated by $\mathrm{B A}^k$, $k=0,1,2,3$, in the charged lepton and neutrino sectors. 
As we have shown, it does not matter to which of these $Z_2$ subgroups $D_4$ is actually broken. It is enough to achieve that
the index $k_l$ in the charged lepton sector is odd, whereas $k_\nu$, the index in the neutrino sector, is even.  
At the same time the hierarchy among charged leptons can be naturally
produced. Apart from $D_4$, responsible for the $\mu-\tau$ symmetric mixing pattern, we employ a cyclic symmetry $Z_5$ in order
to separate the flavons coupling to neutrino and charged lepton sectors at LO as well as to achieve appropriately the
mass hierarchy $m_e \ll m_\mu \ll m_\tau$. Due to a restricted choice of 
flavon fields we find that neutrinos have inverted mass hierarchy with the lightest neutrino mass $m_3 \gtrsim 0.015$ eV. 
These properties together with non-trivial relations among
$m_3$, the observables $|m_{ee}|$ and $\tan\theta_{12}$ as well as the Majorana phases are 
shared with a similar SUSY $D_4$ model, found in \cite{D4SUSY_us}. In the latter, however, the hierarchy among charged leptons
could not be reproduced without fine-tuning. This is due to the fact that right-handed charged leptons also transform as 
$\Rep{1}+\Rep{2}$ under $D_4$. This problem is solved in the present model by assigning them to three
(distinct) one-dimensional representations of $D_4$. The result of $\mu-\tau$ symmetric lepton mixings remains untouched.

A particular feature of our model is that the flavon superpotential not only
leads to the vacuum alignment which is needed in order to arrive at the predictions $\theta_{23}=\pi/4$ and $\theta_{13}=0$,
but also fixes all flavon VEVs through two mass parameters, $M_1$ and $M_2$. In this way, it is natural
to have the same order of magnitude for all flavon VEVs and flat directions in the flavon (super)potential, associated
to free parameters among the VEVs, are avoided. In contrast, in (almost) all models of this type these are generically 
present. The existence of terms with couplings of positive mass dimension also ensures the spontaneous breaking of the flavor symmetry $D_4$ 
and so destabilizes the trivial vacuum in which all flavon VEVs vanish.

We have carefully studied NLO corrections arising from higher-dimensional operators involving several flavons in the Yukawa
sector as well as in the flavon superpotential. The latter induce in general shifts in the flavon VEVs. In the particular case
we discussed here these operators do not disturb at NLO level the LO vacuum structure of the flavons coupling dominantly to the neutrino sector. We eventually
find that all corrections in the neutrino sector are of relative order $\eps^2$. In contrast to this, the results in the charged
lepton sector get corrected at a relative order $\eps$. For this reason, the deviations from $\theta_{23}=\pi/4$ and
$\theta_{13}=0$ of order $\eps$ stem from the charged lepton sector only. This is a further feature which distinguishes the
present model from the one of \cite{D4SUSY_us}, in which deviations from $\mu-\tau$ symmetric lepton mixing are governed by
corrections from the neutrino sector and which at NLO lead to $\theta_{23}-\pi/4$ being much smaller than $\theta_{13}$. 
In the present model NLO corrections are furthermore relevant to generate the electron mass of order $\eps^3 v_d$. As frequently happens in such models,
the mass of the tau lepton is generated through a non-renormalizable operator so that small and moderate values of $\tan\beta$
are preferred.

Several models leading to $\theta_{23}=\pi/4$ and $\theta_{13}=0$ have been considered before in
the literature. The model closest related to the one illustrated is discussed in \cite{D4SUSY_us}. However, as mentioned,
this model needs some fine-tuning to get the correct charged lepton mass hierarchy. The model \cite{GL1}, on which \cite{D4SUSY_us}
is actually based, is in contrast non-SUSY and thus incorporates no solution to the hierarchy problem. Nevertheless, in this model 
it has been shown \cite{GL1,Ksymm_mmu_mtau} that $m_\mu\ll m_\tau$ can arise from a softly broken (additional) symmetry, whereas $m_e$ being 
small can be attributed to a small VEV of one of the Higgs doublets present in the model. A model constructed in the same
spirit as \cite{GL1} can be found in \cite{GLS3} having as flavor group $S_3\times Z_2$ (recall that $S_3$ is isomorphic to
$D_3$). Its results are very similar to those of the model in \cite{GL1}, apart
from fewer constraints on the neutrino mass spectrum. A neat example of a minimalistic SUSY model
leading to $\mu-\tau$ symmetric mixing is given in \cite{S3alt}. 
The flavor group is $S_3\times Z_3$. 
\footnote{In \cite{S3alt} also an extension of the model to the quark sector is discussed, which entails adding another
cyclic symmetry $Z_3^\prime$ and eventually reproduces the quark mass spectra. However, it requires some enhancement in order to
generate a large enough Cabibbo angle.}
The structure of the model is very similar to ours and in the neutrino sector also a $Z_2$ subgroup of the same type as
here is preserved by appropriate flavon VEVs. Yet, in the charged lepton sector an alignment is employed which 
completely breaks $S_3$, but efficiently generates the charged fermion mass hierarchy. The alignment is studied in detail in 
\cite{S3alt}, but, as usual, two parameters among the flavon VEVs remain undetermined, thus
giving rise to flat directions in the flavon (super)potential.

In summary, we have constructed a SUSY $D_4$ model, which predicts $\mu-\tau$ symmetric lepton mixing through breaking $D_4$ to distinct $Z_2$ subgroups
in the charged lepton and neutrino sectors, respectively, and at the same time accommodates naturally the mass hierarchy $m_e\ll m_\mu\ll m_\tau$.
Furthermore, all flavon VEVs are fixed through mass parameters of the superpotential. Thus, 
the problem of free parameters among the flavon VEVs is avoided, which is often
met in models of such type. On the basis of this model it might be very interesting to consider an extension to the quark sector.
As has been shown in \cite{thetaC}, also the Cabibbo angle might arise from a non-trivial breaking of a dihedral group to distinct
$Z_2$ subgroups in up and down quark sectors and so its size might have a similar origin as the prediction of 
$\theta_{23}=\pi/4$ and $\theta_{13}=0$ in the lepton sector.

\subsection*{Acknowledgments}

\noindent We thank Alexander Blum for collaboration at the early stages of this work.
We are grateful to the Galileo Galilei Institute for Theoretical Physics for
hospitality and the INFN for partial support at the beginning of this
work.

\end{document}